\documentclass[journal]{IEEEtran}
\usepackage{cite}
\usepackage{amsmath,amssymb,amsfonts}
\usepackage{algorithmic}
\usepackage{graphicx}
\usepackage{textcomp}
\usepackage{breqn}
\usepackage{amsmath}
\usepackage{multirow}
\usepackage{xparse}
\usepackage{array}
\newcolumntype{M}[1]{>{\centering\arraybackslash}m{#1}}
\newcolumntype{N}{@{}m{0pt}@{}}
\usepackage{stackengine}
\newcommand\xrowht[2][0]{\addstackgap[.5\dimexpr#2\relax]{\vphantom{#1}}}
\NewDocumentCommand{\ceil}{s O{} m}{%
  \IfBooleanTF{#1} 
    {$\left\lceil#3\right\rceil$} 
    {#2\lceil#3#2\rceil} 
}
\def\BibTeX{{\rm B\kern-.05em{\sc i\kern-.025em b}\kern-.08em
    T\kern-.1667em\lower.7ex\hbox{E}\kern-.125emX}}
\begin{document}
\title{Multipath-Enhanced Measurement of Antenna Patterns: Experiment}
\author{Daniel D. Stancil, \IEEEmembership{Life Fellow, IEEE}, and Alexander R. Allen 
\thanks{The authors are with the Department of Electrical and Computer Engineering, North Carolina State University, Raleigh, NC 27695 USA (e-mail: ddstanci@ncsu.edu). }}

\maketitle

\begin{abstract}
In a companion paper we presented the theory for an antenna pattern measuring technique that uses (rather than mitigates) the properties of a multipath environment. Here we use measurements in a typical home garage to experimentally demonstrate the feasibility of the technique. A half-wavelength electric dipole with different orientations was used as both the calibration and test antennas. For simplicity, we limited the modeling of the antenna pattern to using only the three $l=1$ vector spherical harmonics. Three methods were used to analyze the measurements: a matrix inversion method using only 3 sense antennas, a least-square-error technique, and a least-square-error technique with a constant power constraint imposed. The two least-square-error techniques used the measurements from 10 sense antennas. The constrained least-square-error technique was found to give the best results. 
\end{abstract}


\section{Introduction}
Traditional techniques for measuring antenna patterns require minimizing or mitigating the effects of multipath propagation in the environment \cite{hemming_electromagnetic_2002,Xu_anechoic_2019}. In addition to constructing anechoic chambers, example techniques include time-gating to retain only the direct path \cite{loredo_echo_2004}, and averaging out multipath in reverberation chambers using stirrers \cite{fiumara_free-space_2005}. A more complete summary of approaches is given in \cite{stancil_multipath-enhanced_nodate}.

In a companion paper, a technique that \textit{uses} the multipath propagation to obtain pattern information \cite{stancil_multipath-enhanced_nodate} has been introduced. This should enable pattern measurements without the need for costly anechoic environments.

In this work we demonstrate the feasibility of the technique using measurements in a typical home garage. 

\section{Description of Experiment}
\begin{figure}
\centering
\includegraphics[width=0.5\textwidth]{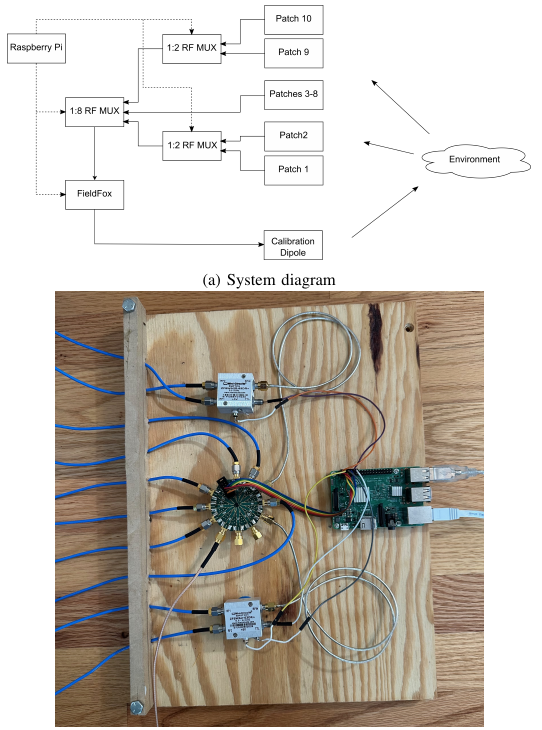}
%
\caption{Top: block diagram of the measurement system. Bottom: photo of the experimental multiplexing system. The interchange of connections to patches 1 and 2 was required owing to the use of a 1:2 switch with reversed outputs.}
    \label{fig:MEAS_sys}      
\end{figure}

To experimentally verify the ideas, we constructed a test system to analyze the case for a half-wave dipole with arbitrary orientation. Based on the discussion in \cite{stancil_multipath-enhanced_nodate}, the system was designed to use up to 10 independent test antenna orientations to form the calibration matrix. An RF measurement system for making the measurements at an operating frequency of 3 GHz is shown in Figure \ref{fig:MEAS_sys}. 
Arbitrarily oriented patch antennas were used for the 10 sense antennas.

The 1:2 multiplexers used are a Minicircuits ZFSWA2R-63DR+ and a ZFSWA2-63DR+ (the difference is that the order of the output connectors is interchanged), and an evaluation board for an Infineon BGSF110GN26 (SP10T switch) was used for the 1:8 multiplexer. (Although the Infineon chip has 10 channels, two of these have low-pass filters below our selected operating frequency and hence could not be used.) RF measurements were made using a Keysight FieldFox\textsuperscript{\textregistered} vector network analyzer under the control of a Raspberry Pi 3B+ microprocessor. 
Crosstalk between channels in the multiplexer system was characterized by connecting the FieldFox output to each patch antenna input in turn and measuring the response at the multiplexer input while stepping through the multiplexer channels. During these measurements, the unexcited patch antenna inputs were terminated in 50 Ohm loads. If $\mathbf{v}$ is the vector of true voltages at the patch antenna inputs, then the measured voltage vector $\mathbf{v}_m$ is given by
\begin{equation}
    \mathbf{v}_m = \mathbf{M}_\times \cdot \mathbf{v},
    \label{eq:crosstalk}
\end{equation}
where $\mathbf{M}_\times$ is the measured cross-talk matrix. It follows that estimates of the true voltages can be recovered with the calculation
\begin{equation}
    \mathbf{v} = \mathbf{M}_\times^{-1} \cdot \mathbf{v}_m.
    \label{eq:crosstalk_cor}
\end{equation}

\begin{figure}
    \centering
    \includegraphics[width=0.4\linewidth]{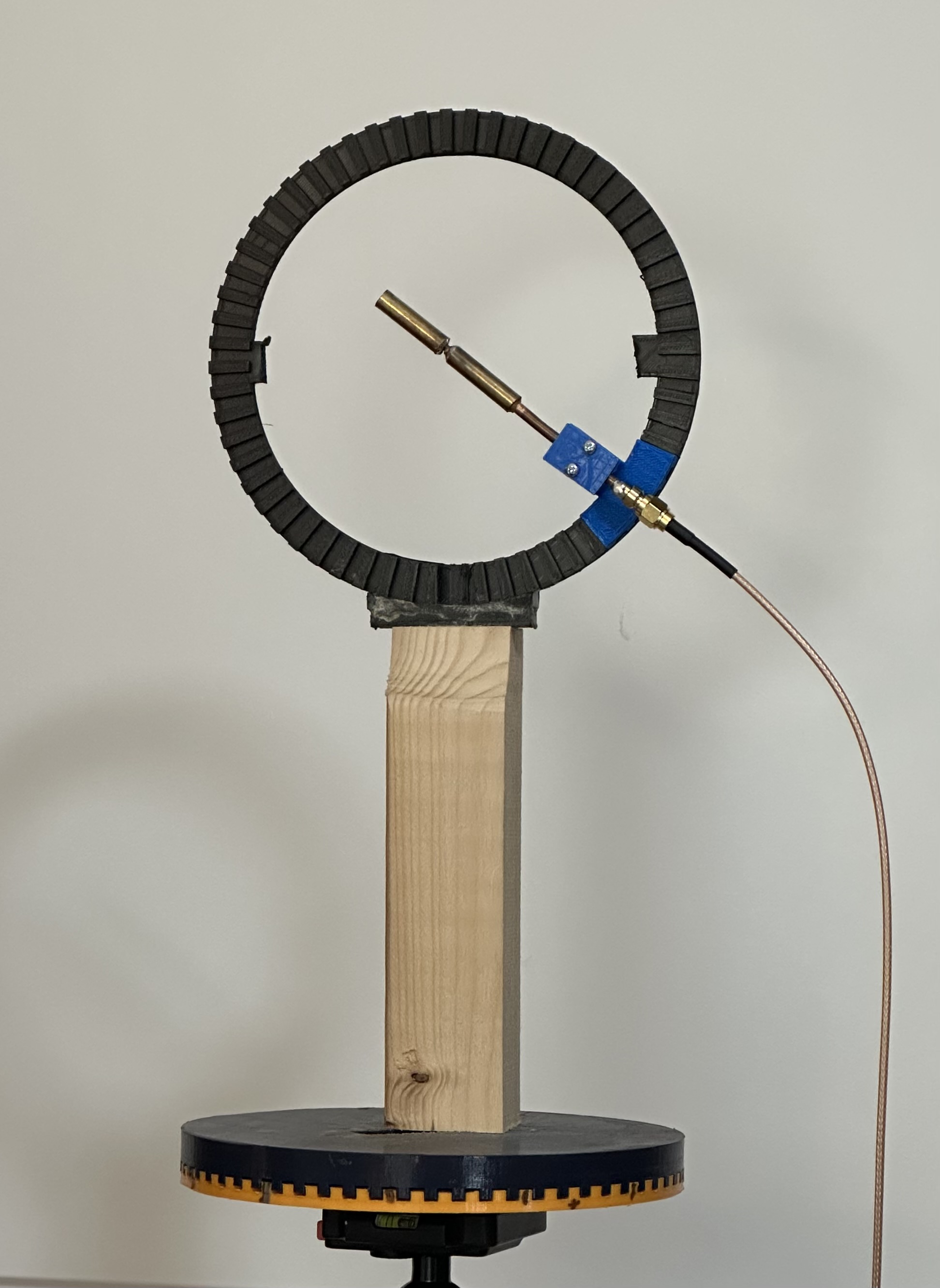}
    \caption{Sleeve dipole mounted on the custom goniometer.}
    \label{fig:goniom}
\end{figure}

\begin{figure}
    \centering
    \includegraphics[width=0.25\linewidth,angle=-90]{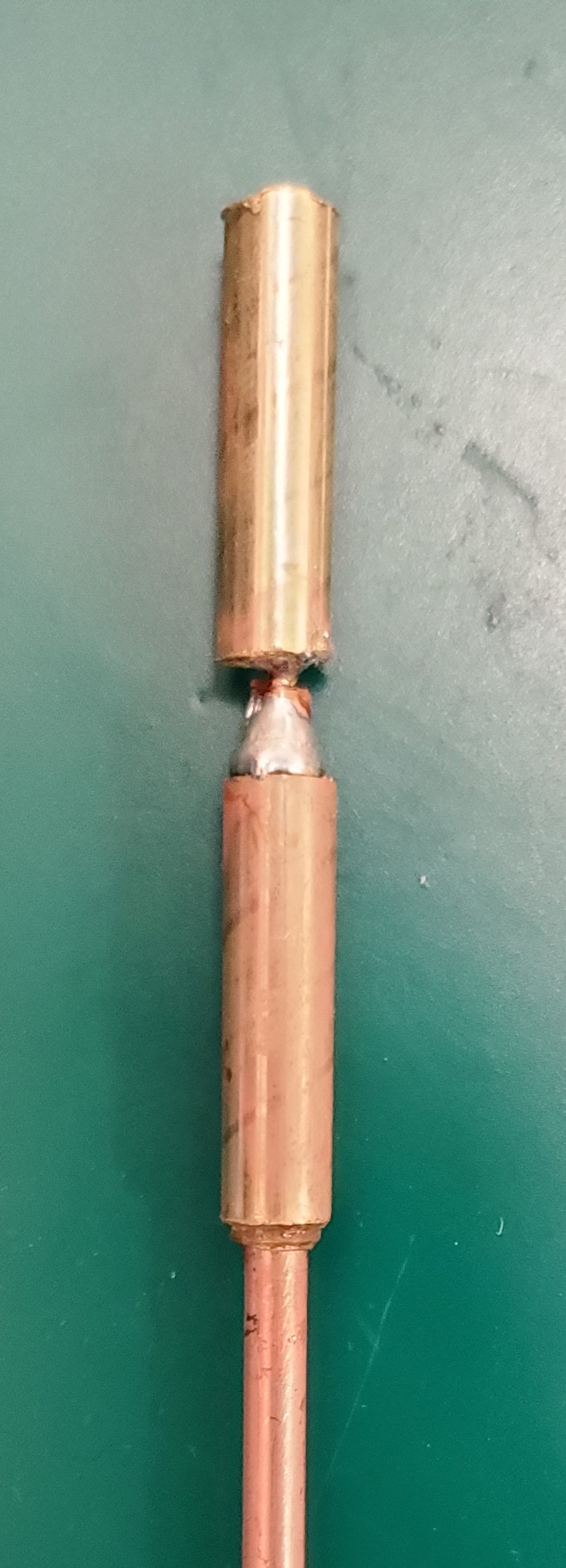}
    \caption{Sleeve dipole used in the experiments.}
    \label{fig:dipole}
\end{figure}

A sleeve dipole with different orientations was used for both calibration as well as the antenna under test. The sleeve dipole mounted on a custom goniometer is shown in Figure \ref{fig:goniom}. The goniometer enabled discrete angular orientations in steps of 6 degrees. A close-up view of the sleeve dipole is shown in Figure \ref{fig:dipole}.

\begin{figure}
    \centering
    \includegraphics[width=0.75\linewidth,angle=-90]{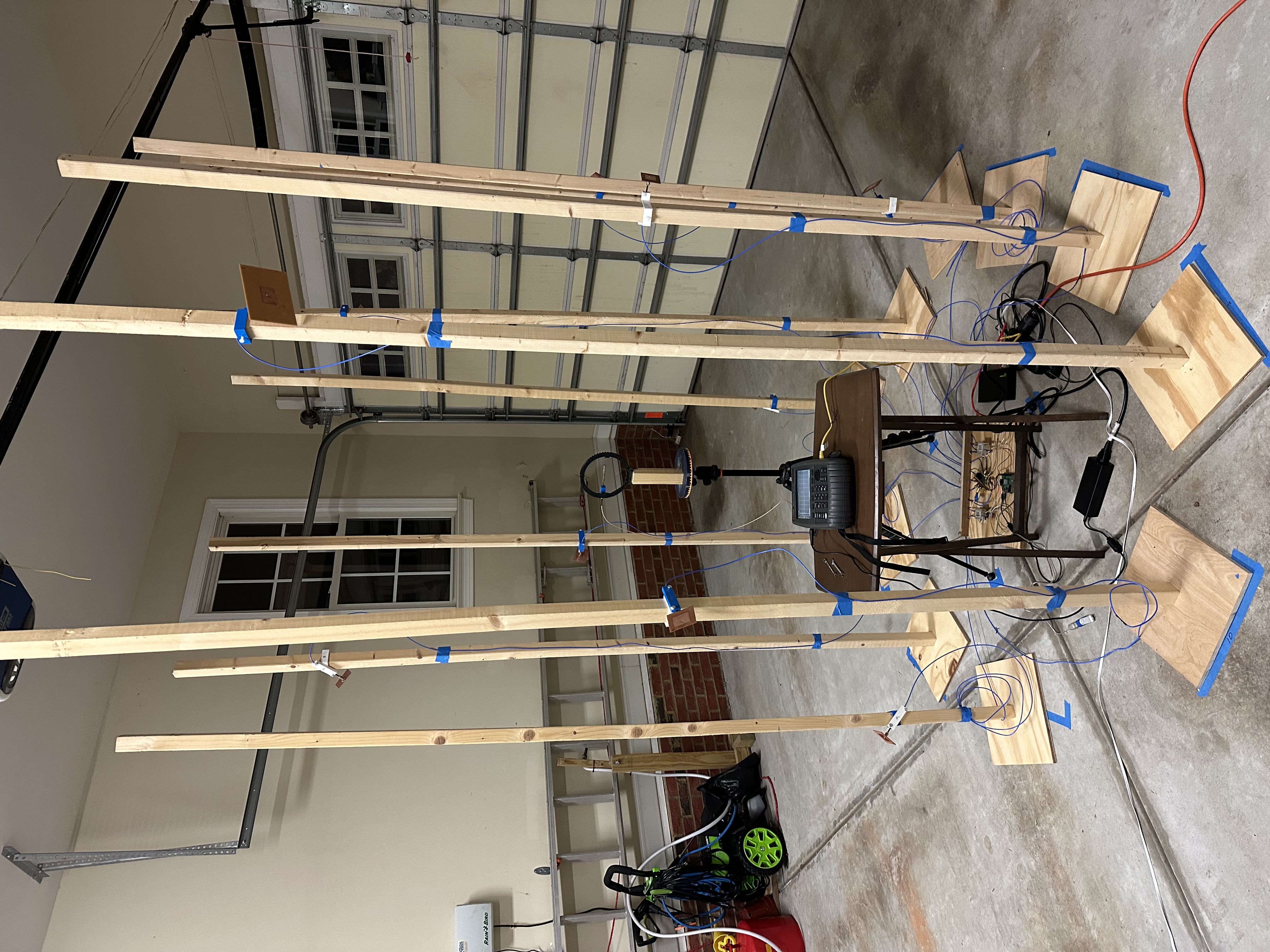}
    \caption{The test setup in a garage. The 10 patch antennas used for sensing are mounted with different heights and orientations on the wooden posts. The test dipole and goniometer are mounted on a camera tripod at the center. The Fieldfox vector network analyzer can be seen on the small table, and the Raspberry Pi, RF switching circuits and ethernet switch are on the floor beneath the tripod. The experiment was initiated from a laptop located outside of the garage and connected to the Raspberry Pi via ethernet. This ensured there was no movement in the environment while the measurements were made.}
    \label{fig:garage}
\end{figure}

\begin{figure}
    \centering
    \includegraphics[width=0.6\linewidth]{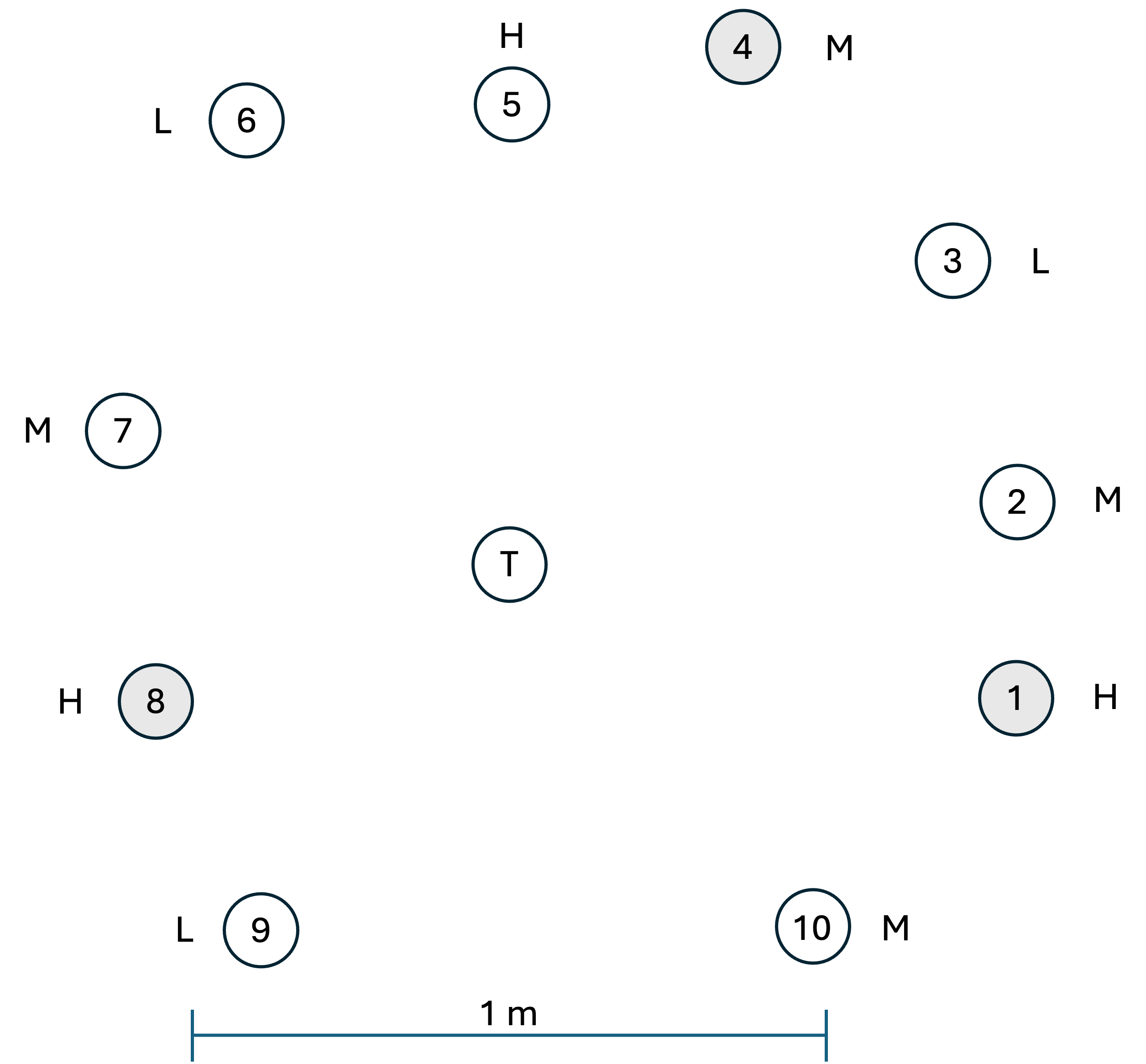}
    \caption{Plan view of approximate layout of the AUT (labeled ``T") and the 10 patch antennas. The patch antennas were located at three different heights above the floor, labeled ``L, M, H". These heights are approximately 0.37m, 1.2m, and 2m, respectively. The AUT was also located at the medium height of about 1.2m. The shaded circles represent the patch antennas that maximized the 1-entropy for the Matrix Inversion method.}
    \label{fig:garage_layout_diag}
\end{figure}

Measurements were made in a garage with a metal garage door, with no steps taken to reduce reflections. The setup is shown in Figure \ref{fig:garage}. The patch antennas can be seen mounted at different heights and orientations on the wooden posts surrounding the test antenna which is mounted on a tripod. 
A plan view approximately to scale (the precise locations were not measured) is shown in Figure \ref{fig:garage_layout_diag}. Other than wanting to represent a variety of heights and directions, there was no pattern to the choice of patch antenna locations.

Ten test antenna orientations were used in anticipation of a full analysis of the electric dipole as discussed above. However, as a first demonstration of the concept, we approximated the dipole with the three strongest harmonics for $l=1$. This only required 3 calibration antennas, which we chose as the dipole oriented along the three orthogonal axes. 

A set of 10 orientations was generated in addition to the calibration antenna orientations to serve as antenna under test (AUT) cases. These are listed in the left columns of Table \ref{tab:summary}. These orientations were selected to represent a reasonably complete coverage of the sphere. Although the reference antennas and the AUTs are realized by different orientations of the same sleeve dipole, we will frequently refer to these as if they were separate antennas for convenience (e.g., ``antenna reference orientations" may be referred to interchangeably as ``reference antennas").

\begin{table*}[]
    \caption{Summary of experimental results}
    \label{tab:summary}
    \centering
    \begin{tabular}{|c|c|c|c|c|c|c|c|c|N}
    \hline\xrowht[()]{5pt}
    \multirow{2}{*}{Orientation \#}& \multirow{2}{*}{$\theta$  (deg)}  & \multirow{2}{*}{$\phi$ (deg)} &\multicolumn{2}{|c|}{Matrix Inverse}&\multicolumn{2}{|c|}{LSE}&\multicolumn{2}{|c|}{Constrained LSE}\\
    \cline{4-9}\xrowht[()]{5pt}
    \multirow{2}{*}{}&  &  & RMSE (dB)& $R_r$ $(\Omega)$& RMSE (dB)&$R_r$ $(\Omega)$ &RMSE (dB) &$R_r$ $(\Omega)$\\
    \hline
    1 &  18    &     0 & -20.3786 &  55.6393  &-20.5676 &  54.3609 & -23.2998 &  72.9016\\
    \hline
    2 &  18 & 120 & -18.1382  & 74.5371  &-26.4792 &  61.5702 & -35.7840 &  72.9016\\
    \hline
    3 &  18 &-120 & -16.0723 &  43.3391  &-13.8468  & 37.5476 & -17.4519 &  72.9016\\
    \hline
    4 &  42 &  60 & -11.0097 &  46.2895  &-16.2261 &  49.6765 & -16.4016 &  72.9015\\
    \hline
    5 &  42 & 180 & -15.3328  & 55.8037  &-13.8474 &  37.2317 & -19.5179 &  72.9016\\
    \hline
    6  & 42 & -60 & -10.7356 &  42.6464  &-13.6183 &  43.4529 & -12.5105 &  72.9016\\
    \hline
    7 &  90 &  30 & -21.0854 &  79.8493  &-19.7381 &  74.7897 & -19.9634 &  72.9016\\
    \hline
    8 &  90 & 120 & -15.7867 &  54.2604  &-14.6552 &  35.2741&  -19.8868 &  72.9016\\
    \hline
    9 &  90 & 210 & -15.4119 &  37.5729  &-16.3295 &  54.1650 & -16.3971  & 72.9016\\
    \hline
   10  & 90&  -60 &  -7.8009  & 86.9811  &-14.8634 &  58.8408&  -14.0171 &  72.9016\\
   \hline
   mean values& & & -15.1752& 57.6919& -17.0172& 50.6909 &-19.5230& 72.9016\\
   \hline
    \end{tabular}
\end{table*}

The sensing/patch antenna locations used three sets of heights: 3 above the test antenna, 3 below the test antenna, and 4 approximately at the same height as the test antenna. The patch mounting brackets could be set so that the direction perpendicular to the plane of the patch is either in the horizontal direction, or at angles of ±45 degrees from horizontal. The angle was initially chosen for each patch antenna that most nearly directed the main lobe of the patch antenna pattern toward the test antenna. However, to emphasize the role of multipath, the wooden posts were then rotated by about 180 degrees so that the patch antennas were outward facing instead of pointing toward the antenna.

\section{Analysis Methods}
Three methods of analysis were used, as described in \cite{stancil_multipath-enhanced_nodate}. We will refer to these as the Matrix Inversion (MI) method, the Least Square Error (LSE) method, and the Constrained Least Square Error (CLSE) method.

The general approach was to first take measurements of the sense antennas for each of a set of reference or calibration antennas, then take measurements on the antennas under test (AUT), and finally to determine the  superpositions of the reference antenna signals that best approximate the  of measurements of the AUTs. Since the vector spherical harmonic (VSH) composition of the reference antennas is assumed to be known, the weighted sum of these yields an approximation to the VSH components of the AUT, from which the pattern can be constructed. 

As discussed in \cite{stancil_multipath-enhanced_nodate}, the key properties of the channel are represented by the matrix $\mathbf{T}=\mathbf{V}_R\cdot\mathbf{A}_R^{-1}$, where $\mathbf{V}_R$ is the matrix containing the patch antenna measured voltages for the calibration antennas, and $\mathbf{A}_R$ is the matrix containing the VSH coefficients for the calibration antennas. Consequently obtaining the best result requires optimizing the choice of calibration antennas as well as the placement of the sense antennas (patches in the present case). However, since the matrix $\mathbf{A}_R$ is well conditioned in this case, we can consider $\mathbf{V}_R$ to represent the channel. We proceed by separating the process into two steps: the first is determining the weights of the reference antennas needed to construct the patch antenna voltages from the AUT, and the second is calculating the resulting VSH coefficients and patterns. The accuracy of this second step depends on the accuracy with which the VSH coefficients for the reference antennas are known. Measuring or calculating the coefficients of the reference antennas is an involved process on its own, so for illustration purposes, we will assume the VSH coefficients of our sleeve dipole at various orientations are the same as those of a thin half-wave dipole, as computed in \cite{stancil_multipath-enhanced_nodate}. The weight vector $\mathbf{w}$ associated with a particular AUT is related to the measured voltages at the patch antenna as
\begin{equation}
    \mathbf{v} = \mathbf{V}_R\cdot\mathbf{w}.
\end{equation}
Because of the linearity of the system, this weight vector also determines the VSH coefficient vector $\mathbf{a}$ of the AUT:
\begin{equation}
    \mathbf{a}=\mathbf{A}_R\cdot\mathbf{w}.
    \label{eq:a_calc}
\end{equation}

Also, since the AUTs are represented simply by rotations, it is possible to write down the exact solution for the weights to compare with the experimental analysis:
\begin{align}
    \mathbf{w}_n &= \left[w_{z,n}, w_{x,n}, w_{y,n}\right]^T\nonumber\\
    &=\left[\cos\theta_n,\sin\theta_n\cos\phi_n,\sin\theta_n\sin\phi_n\right]^T,
\end{align}
where $\mathbf{w}_n$ is the weight vector corresponding to the orientation $(\theta_n,\phi_n)$, with the angles defined as in Fig. 2 of \cite{stancil_multipath-enhanced_nodate}.

In the following sections we obtain estimates of the weight vector $\mathbf{w}^e$ using three methods. To evaluate the accuracy of our estimates, we introduce the weight rms error:
\begin{align}
    \Delta w_{RMS} &= \sqrt{(\mathbf{w}^e-\mathbf{w}_n)^T\cdot (\mathbf{w}^e-\mathbf{w}_n)/N},\\
    RMSE\, (dB) &= 20\log_{10}(\Delta w_{RMS}).
    \label{eq:w_RMS}
\end{align}
with $N=3$ in our present case.

Once estimates of the weights are obtained, the VSH coefficient vectors are obtained from (\ref{eq:a_calc}).
The radiated power is then
\begin{equation}
    P = \frac{|\mathbf{a^e}|^2}{2\eta_0 k^2},
\end{equation}
where $\eta_0$ is the impedance of free space, and $k=2\pi/\lambda$ is the wave number. The radiation resistance can be obtained from the radiated power:
\begin{equation}
    R_r = \frac{2P}{|I|^2},
\end{equation}
where $I$ is the current at the terminals of the antenna.

\subsection{Matrix Inversion Method}
Measurements of signals from the three calibration antennas were made at all 10 patch antennas. However, to obtain square matrices, 3 of the 10 must be selected, resulting in 120 possible choices. This was done by choosing the set of 3 patches that resulted in the maximum 1-entropy $H_1$ \cite{behjoo_optimal_2022}
\begin{equation}
    H_1 (\mathbf{V}_R) = \frac{1}{2}\log_2\left(\det{(\mathbf{V}_R\cdot\mathbf{V}_R^\dagger)}\right),
    \label{eq:entropy}
\end{equation}
where the $\dagger$ superscript represents the congugate transpose. The weight estimates are then obtained from
\begin{equation}
    \mathbf{w}^e = \mathbf{V}_R^{-1}\cdot\mathbf{v}.
    \label{eq:w_estimate}
\end{equation}

\begin{figure}
    \centering
    \includegraphics[width=\linewidth]{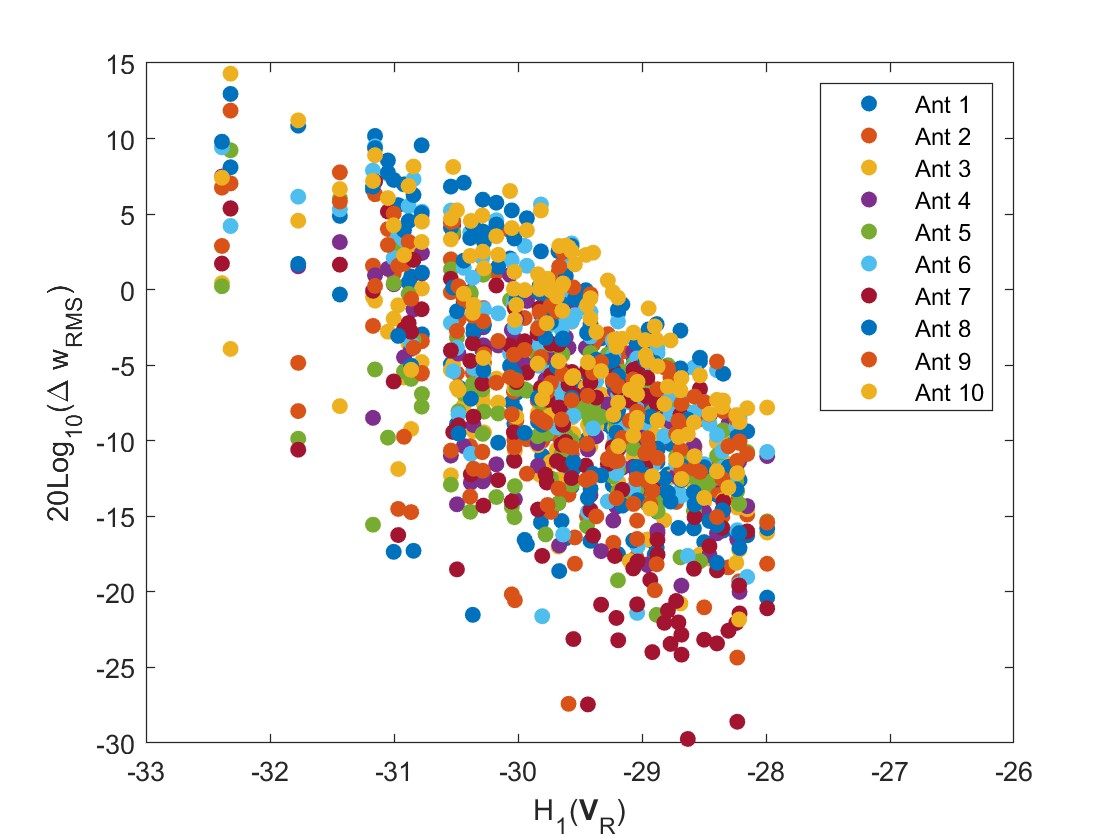}
    \caption{RMS error in the weights $\mathbf{w}$ versus the 1-entropy $H_1$.}
    \label{fig:RMSEvsH1}
\end{figure}

\begin{figure}
    \centering
    \includegraphics[width=\linewidth]{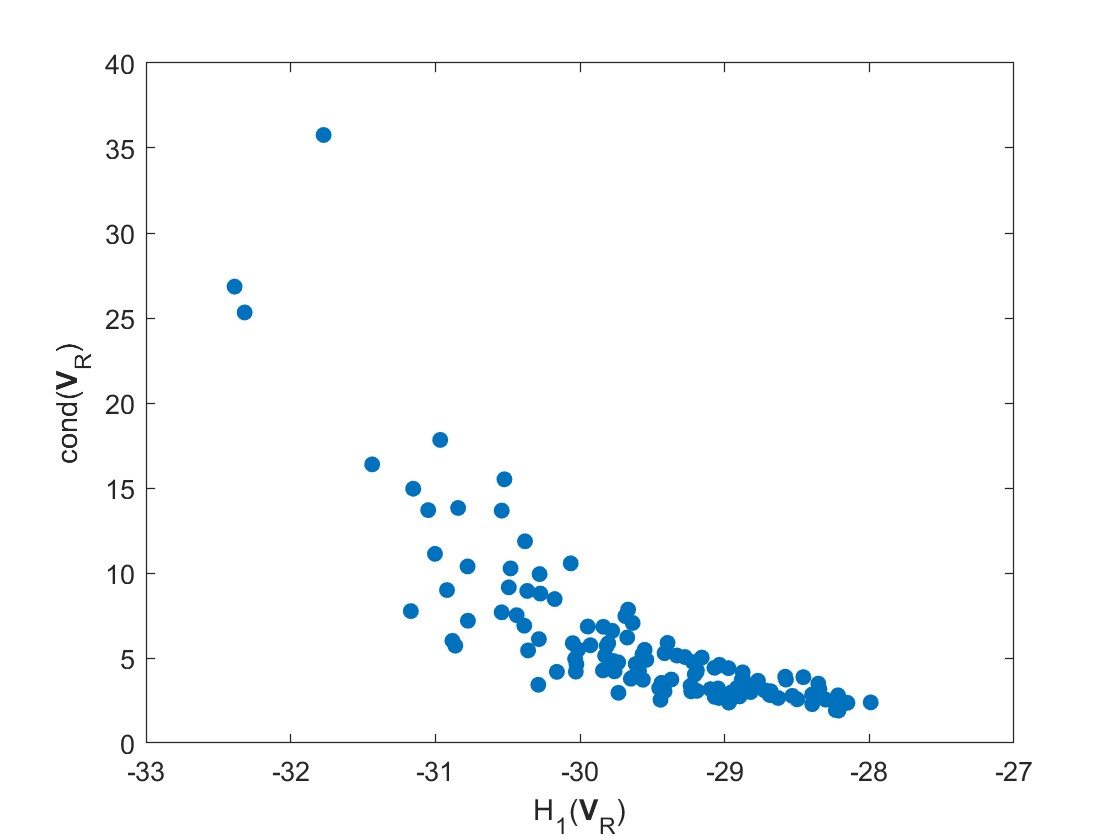}
    \caption{Comparison between the condition of the channel matrix versus its 1-entropy $H_1$.}
    \label{fig:CONDvsH1}
\end{figure}

To verify that $H_1$ is indeed a useful metric, each of the 10 AUTs were evaluated for each of the 120 possible choices for $\mathbf{V}_R$ by calculating the rms error of the weight vector using (\ref{eq:w_RMS}). The results are shown in Figure \ref{fig:RMSEvsH1}. Although there is significant scatter in the data, the trend of minimizing the error by maximizing the 1-entropy is clear. Another potential metric mentioned in \cite{stancil_multipath-enhanced_nodate} is cond($\mathbf{V}_R$). A comparison between the matrix condition and the 1-entropy is shown in Figure \ref{fig:CONDvsH1}. There is a clear correlation in that minimizing the condition number tends to maximize the 1-entropy. However, again there is statistical spread, and in fact the two metrics give different optimum patch antenna choices in our case. In our analysis, the set of patch antennas that maximized the 1-entropy was antennas 1, 4, and 8, as shown by the shaded circles in Figure \ref{fig:garage_layout_diag}. This set of antennas was used to calculate the weight estimates using Eq. (\ref{eq:w_estimate}). The results are shown in Table \ref{tab:summary}. 

Also shown in the table are the resulting values of radiation resistance $R_r$ obtained from the weights, assuming $\mathbf{A}_R$ corresponds to thin half-wave dipoles. The theoretical value is $R_r = 72.9\,\Omega$, if only the $l=1$ VSH terms are included. The deviations from this value indicate errors in the amplitude of the reconstructed pattern. In addition to errors in pattern amplitude, orientation errors can occur as well. These errors become apparent when the pattern is plotted.

\subsection{Least-Square-Error Method}

Unlike the Matrix Inverse method, the Least-Square-Error (LSE) method can make use of any number of sense antennas (in our case, patch antennas) that is greater than the number of needed VSH coefficients. From \cite{stancil_multipath-enhanced_nodate}, the LSE weight vector estimate is given by
\begin{equation}
  \mathbf{w} = \left[\text{Re}\{{\mathbf{V}_R}^\dagger\cdot \mathbf{V}_R\}\right]^{-1}\cdot \text{Re}\{{\mathbf{V}}_R^\dagger\cdot {\mathbf{{v}}}\}.
\end{equation}
Here $\mathbf{V}_R$ is a $10\times3$ matrix. The RMSE for each AUT along with the calculated radiation resistance are also given in Table \ref{tab:summary}.

\subsection{Constrained Least-Square-Error Method}
The RMSE errors can be improved by imposing a constraint of constant power. For our calculations we used the form
\begin{equation}    
    \mathbf{w}^{\text{T}}\cdot\mathbf{w} = 1.
\end{equation}
The error minimization subject to this constraint was implemented with the MATLAB\textsuperscript{\textregistered} function {\fontfamily{pcr}\selectfont
fmincon} using the {\fontfamily{pcr}\selectfont
sqp} algorithm. The equality constraint used was
\begin{equation}
    \verb|ceq = w'*w - 1;|
\end{equation}
As can be seen in Table \ref{tab:summary}, for the present case this also ensured that the radiation resistance was equal to the theoretical value for the assumed calibration antennas. This is a consequence of the use of orthogonal calibration antennas.

\section{Discussion}
From Table \ref{tab:summary}, we see that the constrained LSE method generally gives the best result, while the Matrix Inversion method gives the least accurate. Although not a problem in this special example, the MI method would also have problems with ill-conditioned matrices if some of the SVH coefficients were zero or extremely small \cite{stancil_multipath-enhanced_nodate}. This would likely often be the case when there is very little a priori knowledge about the VSH spectrum for the AUT.

\begin{figure}
\includegraphics[width=0.5\textwidth]{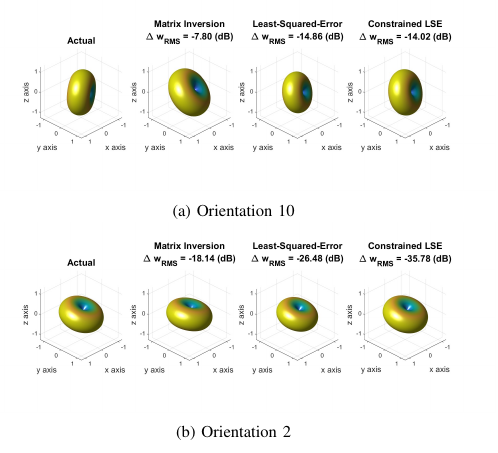}
      \caption{Top: Patterns from one of the AUTs with the largest error (Orientation 10). Bottom: Patterns from one of the AUTs with the smallest error (Orientation 2). In both cases the left-most pattern is the theoretical pattern for the known orientation.}
    \label{fig:patterns}      
\end{figure}

As mentioned previously, error contributions to RMSE (dB) come from both the amplitude and the orientation of the reconstructed pattern. The orientation errors can be best seen when the patterns are plotted (Figure \ref{fig:patterns}). 


From Table \ref{tab:summary}, we see that Orientation 10 had some of the largest errors. The patterns corresponding to this case are shown in Figure \ref{fig:patterns}(a). Errors in orientation are apparent in all three cases, and errors in amplitude can be seen in the MI and LSE cases.

In contrast, Orientation 2 had some of the lowest errors. Referring to Figure \ref{fig:patterns}(b), while some errors are apparent in the MI and LSE cases, the Constrained LSE pattern is almost indistinguishable from the actual pattern.

Avenues to improve the accuracy of the measurements include the use of a precision goniometer, use of 1:$N$ multiplexers with minimal crosstalk, and better vector network analyzer calibration (calibration using the short-open-load-through (SOLT) procedure was performed, but the impedance standards used were not precision laboratory standards).

It should also be possible to significantly improve these results by introducing more sources of multipath in the environment, and optimizing the placement of the sense antennas. This can be quantified by organizing the environment to maximize the 1-entropy.  In the more general case where inclusion of VSHs with $l>1$ are needed, it will also be important to design calibration antennas to maximize the 1-entropy with $\mathbf{V}_R\cdot\mathbf{A}_R^{-1}$ serving as the channel matrix in (\ref{eq:entropy}) instead of $\mathbf{V}_R$.

\section{Summary and Conclusion}
In contrast to conventional antenna pattern measurement methods that seek to minimize the effects of multipath propagation, we have demonstrated a technique that uses  multipath propagation to construct the pattern. The approach can be considered similar to a MIMO channel where the inputs are the vector spherical harmonics of the antenna under test, and the outputs are the measured voltages at sense antennas distributed in the multipath environment. By characterizing the environment with a set of calibration antennas, the vector spherical harmonics of the antenna under test can be calculated from the measured sense antenna voltages. The pattern can then be constructed from the vector spherical harmonics. Because the required number of vector spherical harmonics increases rapidly with antenna size, it is likely that this method will be of most interest for characterizing small antennas (of order one wavelength or smaller) such as those typically used in consumer devices.



\bibliographystyle{IEEEtran}
\bibliography{IEEEabrv,myzotero3}

\begin{thebibliography}{1}
\providecommand{\url}[1]{#1}
\csname url@samestyle\endcsname
\providecommand{\newblock}{\relax}
\providecommand{\bibinfo}[2]{#2}
\providecommand{\BIBentrySTDinterwordspacing}{\spaceskip=0pt\relax}
\providecommand{\BIBentryALTinterwordstretchfactor}{4}
\providecommand{\BIBentryALTinterwordspacing}{\spaceskip=\fontdimen2\font plus
\BIBentryALTinterwordstretchfactor\fontdimen3\font minus \fontdimen4\font\relax}
\providecommand{\BIBforeignlanguage}[2]{{%
\expandafter\ifx\csname l@#1\endcsname\relax
\typeout{** WARNING: IEEEtran.bst: No hyphenation pattern has been}%
\typeout{** loaded for the language `#1'. Using the pattern for}%
\typeout{** the default language instead.}%
\else
\language=\csname l@#1\endcsname
\fi
#2}}
\providecommand{\BIBdecl}{\relax}
\BIBdecl

\bibitem{hemming_electromagnetic_2002}
L.~H. Hemming, \emph{Electromagnetic {Anechoic} {Chambers}}.\hskip 1em plus 0.5em minus 0.4em\relax IEEE Press, Wiley-Interscience, 2002.

\bibitem{Xu_anechoic_2019}
Q.~Xu and Y.~Huang, \emph{Anechoic and Reverberation Chambers: Theory, Design, and Measurements}.\hskip 1em plus 0.5em minus 0.4em\relax IEEE Press, Wiley-Interscience, 2019.

\bibitem{loredo_echo_2004}
\BIBentryALTinterwordspacing
S.~Loredo, M.~Pino, F.~Las-Heras, and T.~Sarkar, ``Echo identification and cancellation techniques for antenna measurement in non-anechoic test sites,'' \emph{IEEE Antennas and Propagation Magazine}, vol.~46, no.~1, pp. 100--107, Feb. 2004. [Online]. Available: \url{https://ieeexplore.ieee.org/abstract/document/1296154}
\BIBentrySTDinterwordspacing

\bibitem{fiumara_free-space_2005}
\BIBentryALTinterwordspacing
V.~Fiumara, A.~Fusco, V.~Matta, and I.~Pinto, ``Free-space antenna field/pattern retrieval in reverberation environments,'' \emph{IEEE Antennas and Wireless Propagation Letters}, vol.~4, pp. 329--332, 2005. [Online]. Available: \url{https://ieeexplore.ieee.org/document/1512069}
\BIBentrySTDinterwordspacing

\bibitem{stancil_multipath-enhanced_nodate}
D.~D. Stancil, ``Multipath-{Enhanced} {Measurement} of {Antenna} {Patterns}: {Theory},'' \emph{IEEE Trans. Antennas Prop.}, to appear.

\bibitem{behjoo_optimal_2022}
\BIBentryALTinterwordspacing
H.~R. Behjoo, A.~Pirhadi, and R.~Asvadi, ``Optimal {Sampling} in {Spherical} {Near}-{Field} {Antenna} {Measurements} by {Utilizing} the {Information} {Content} of {Spherical} {Wave} {Harmonics},'' \emph{IEEE Transactions on Antennas and Propagation}, vol.~70, no.~5, pp. 3762--3771, May 2022. [Online]. Available: \url{https://ieeexplore.ieee.org/abstract/document/9664490?}
\BIBentrySTDinterwordspacing

\end{thebibliography}

\begin{IEEEbiography}[{\includegraphics[width=1in,height=1.25in,clip,keepaspectratio]{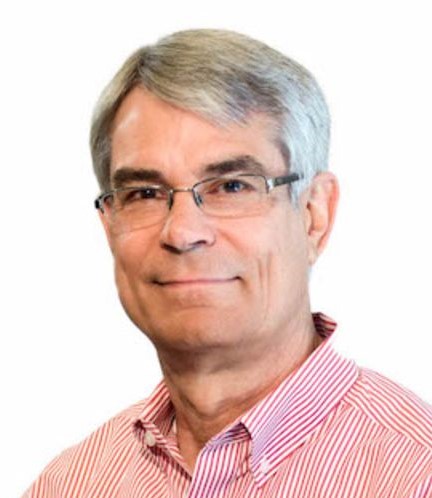}}]{Daniel D. Stancil} (S'75--M'81--SM'91--F'04--LF'20) received the B.S. degree in electrical engineering from Tennessee Technological University, Cookeville, in 1976 and the S.M., E.E., and Ph.D. degrees in electrical engineering from the Massachusetts Institute of Technology (MIT), Cambridge, in 1978, 1979, and 1981, respectively.  

From 1981 to 1986, he was an Assistant Professor of Electrical and Computer Engineering at North Carolina State University. From 1986 to 2009, he was an Associate Professor, then Professor of Electrical and Computer Engineering at Carnegie Mellon University, Pittsburgh, PA. At CMU he served as Associate Department Head of the ECE Department from '92-'94, and Associate Dean for Academic Affairs in the College of Engineering from '96-'00. He returned to North Carolina State and served as Head of the Electrical and Computer Engineering Department from 2009-2023. From 2019-2024 he was the Executive Director of the IBM Q Hub at NC State, a collaboration with IBM on quantum computing. He became Alcoa Distinguished Professor Emeritus in 2024. His research has included such varied topics as spin waves, optics, microwaves, wireless channels, antennas, remote labs, and particle physics.  

Technology for distributing wireless signals through HVAC ducts that Dr. Stancil and his students developed has been installed in such major buildings as Chicago's Trump Towers and McCormick Place Convention Center. The demonstration of neutrino communications by a multidisciplinary team coordinated by Dr. Stancil was recognized by Physics World Magazine as one of the top 10 Physics Breakthroughs of 2012. Additional recognitions that his work has received have included an IR 100 Award and a Photonics Circle of Excellence Award. Dr. Stancil has served as president of the IEEE Magnetics Society and president of the ECE Department Heads Association.
\end{IEEEbiography}

\begin{IEEEbiography}[{\includegraphics[width=1in,height=1.25in,clip,keepaspectratio]{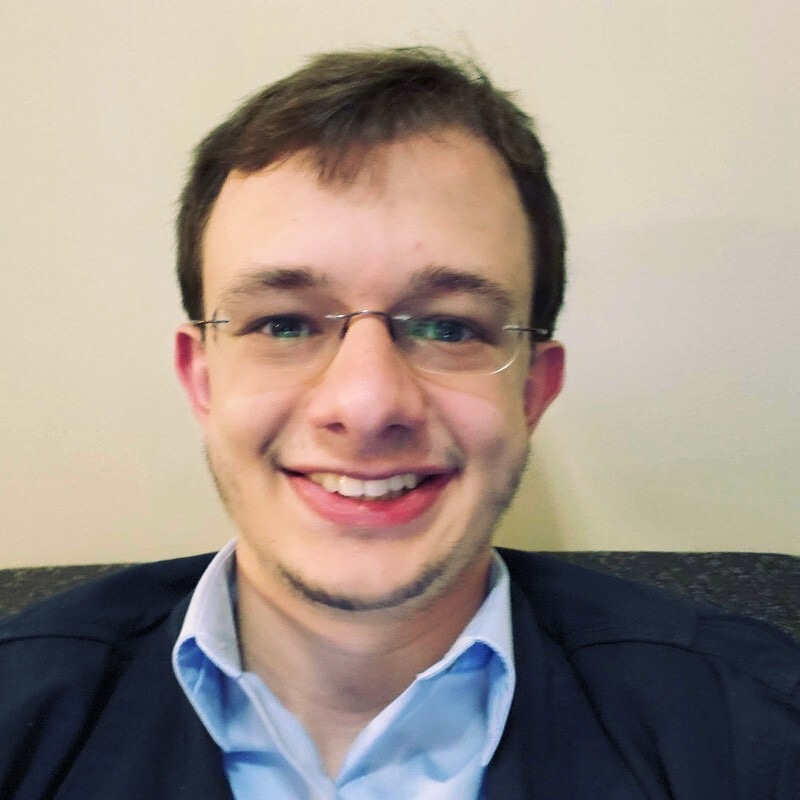}}]{Alexander R. Allen} received the B.S. degree in Electrical Engineering and Computer Engineering from NC State University in 2021. From 2020 to 2025 he was a software engineer at NVIDIA working on kernel mode device drivers for network switches and display drivers for SoCs. He is currently serving as Director of System Software Engineering for Precision Sustainable Agriculture, a USDA research network composed of 32 universities and 5 grower councils, where he leads the development of high-throughput plant phenotyping software for improving crop sustainability.
\end{IEEEbiography}

\end{document}